\documentclass[12pt]{article}
\usepackage{fullpage}
\usepackage{amsmath}
\usepackage{amssymb}
\usepackage[]{graphicx}
\graphicspath{{graphics/}} 

\usepackage{epstopdf}
\usepackage{multirow}
\usepackage{color}
\usepackage[linktocpage]{hyperref}
\usepackage{cite}
\usepackage{wrapfig}
\usepackage{subeqnarray}
\usepackage{lscape}
\usepackage[mathscr]{eucal}
\usepackage{dcolumn}
\newcolumntype{d}[1]{D{.}{\cdot}{#1} }

\usepackage{upgreek}

\def\blue{\textcolor{black}}

\def\##1{{\bf #1}}
\def\=#1{\underline{\underline #1}}
\def\ux{\#u_x}
\def\uy{\#u_y}
\def\uz{\#u_z}

\def\tq{\tilde{q}}

\def\le{\left(}
\def\ri{\right)}

\def\lec{\left\{}
\def\ric{\right\}}

\def\lambdao{\lambda_{\scriptscriptstyle 0}}
\def\ko{k_{\scriptscriptstyle 0}}

\def\propdist{\Delta_{prop}}
\def\req{Re($\tq$)}
\def\imq{Im($\tq$)}
\def\gra{~$^\circ$C}

\def\fref#1{Fig.~\ref{#1}}

\def\sref#1{Sec.~\ref{#1}}
\def\tref#1{Table~\ref{#1}}

\def\quadr#1{\left[#1\right]}

\def\quote#1{\textquotedblleft #1\textquotedblright}

\begin{document}
\large
{\bf Signatures of thermal hysteresis in Tamm-wave propagation}

\normalsize

\vspace{10pt}

Francesco Chiadini$^{1}$,
Vincenzo Fiumara$^{2}$,
Tom G. Mackay$^{3,4}$,
Antonio Scaglione$^{1}$, and
Akhlesh Lakhtakia$^{4}$

\vspace{10pt}

$^{1}$Department of Industrial Engineering,
	University of Salerno, via Giovanni Paolo II, 132 -- Fisciano (SA),
	84084, Italy;
	
	 $^{2}$School of Engineering, University of
	Basilicata, Viale
	dell'Ateneo Lucano 10, 85100 Potenza, Italy;
	
$^{3}$School of Mathematics and
   Maxwell Institute for Mathematical Sciences,
University of Edinburgh, Edinburgh EH9 3FD, UK;

 $^{4}$Department of Engineering Science and Mechanics, Pennsylvania State University,
	University Park, PA 16802--6812,
	USA

\vspace{10pt}

\begin{abstract}
We numerically solved the boundary-value problem for Tamm  waves  (which may also be classified as Uller--Zenneck waves here)
guided by the  {planar} interface of a {homogeneous  isotropic dissipative dielectric} (HIDD) material 
and a periodically multilayered isotropic dielectric material.
The HIDD material  was chosen to be VO${}_2$ which, at optical wavelengths,  
has a temperature-dependent refractive index with a hysteresis feature, i.e., the temperature-dependence of its refractive index varies depending  upon whether the temperature is increasing or decreasing.
A numerical code was  implemented  to extract  solutions of the   dispersion equation at a fixed wavelength for both $p$-  and $s$-polarization states over the temperature range  $\quadr{50,80}$\gra .
A multitude of Tamm waves  of both  linear polarization states were found,  demonstrating a  clear demarcation of the heating and cooling phases   in terms of 
wavenumbers and propagation distances.  Thereby, the signatures of thermal hysteresis in Tamm-wave propagation were revealed.
\end{abstract}


\section{Introduction} \label{sec:intro}
An electromagnetic surface wave (ESW) can be guided by the planar interface  of two different materials~\cite{Boardman,PMLbook}. Various types of ESWs can occur whose  classifications and characteristics are determined by the materials forming the interface~\cite{ChJOSAB}.  Included in these types are  Tamm waves and  Uller--Zenneck waves.

 Tamm waves are ESWs guided by the planar interface of two isotropic dielectric materials, one of them being periodically inhomogeneous in the direction normal to the interface. The existence of Tamm waves has been predicted theoretically~\cite{YYH} and  {established} experimentally~\cite{YYC}, and  these ESWs have been harnessed for optical-sensing applications~\cite{SR2005,KA2007,Sinibaldi,KKAVSD}.

Uller-Zenneck waves are ESWs guided by the planar interface of two dielectric materials, one of which is required to be dissipative. 
The existence of Uller-Zenneck waves was predicted theoretically in the early years of the twentieth century~\cite{Uller,Zenneck,Sommerfeld1,Sommerfeld2,Sommerfeld3}, but  unambiguous experimental confirmation was only provided very recently~\cite{FLol}.

In this paper we solve the canonical boundary-value problem of ESWs guided by the planar interface of two isotropic dielectric materials, one of which is periodically inhomogeneous in the direction normal to the interface and the other is a dissipative material.  The ESWs under consideration may be classified as either  Tamm waves or  Uller--Zenneck waves; for definiteness,  the former classification is chosen here.

The  numbers and propagation characteristics of such ESWs are   determined by the structural and constitutive properties of the two partnering materials on either side of the interface. As an example, the periodicity of one of the two partnering materials may determine the number of the ESWs that can be excited~\cite{ChJNP17}. Furthermore, if a  partnering material has  temperature-sensitive constitutive properties, 
then the number and propagation characteristics of ESWs may be controlled by varying the temperature; indeed,  Dyakonov surface waves may be converted to surface--plasmon--polariton waves by increasing the temperature, for example~\cite{ChJOPT17}.

For the partnering material that is a  homogeneous isotropic dissipative dielectric (HIDD) material we chose  VO$_2$ 
 which, at optical wavelengths,  
has a temperature-dependent refractive index $n_d$ \cite{Corm} with a hysteresis \cite{hysteresis} feature, i.e., the temperature-dependence of $n_d$ varies depending  upon whether the temperature is increasing or decreasing.
For
the periodically nonhomogeneous partnering material, we chose a 
periodic multilayered
isotropic dielectric (PMLID)  material, comprising layers of alloys of silicon oxide and silicon nitride, whose refractive indexes were assumed to be insensitive to temperature over the
 range $\quadr{50,80}$\gra\  for the  near-infrared   regime. The PMLID material is nondissipative.

The goal of our study is to delineate the signatures of thermal hysteresis in  Tamm-wave propagation.
The paper is organized as follows: in \sref{sec:matmeth} the canonical boundary-value problem for ESW  propagation guided by the  planar HIDD/PMLID  interface is presented.
In \sref{sec:RaD} numerical results highlighting the signatures of thermal hysteresis in  Tamm-wave propagation, 
at a fixed  free-space  wavelength   $\lambdao=800$~nm, are reported and discussed   for both $p$- and $s$-polarization states.  Conclusions follow in Sec.~\ref{sec:cr}.

\section{Theoretical Preliminaries and Materials}\label{sec:matmeth}
\begin{figure}
	\begin{center}
		\begin{tabular}{c}
			\includegraphics[width=0.95\linewidth]{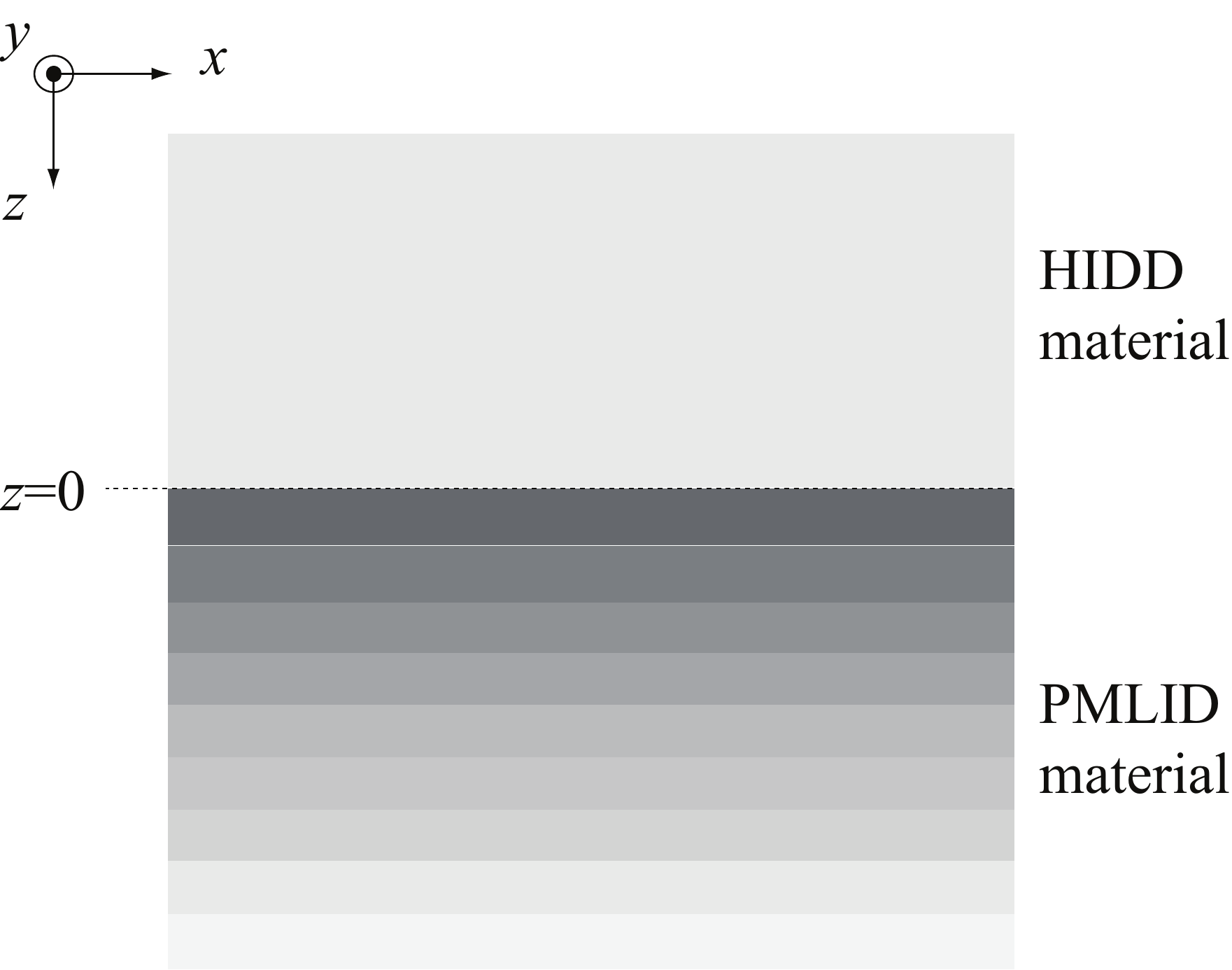}
		\end{tabular}
	\end{center}
	\caption[Schematic]{ Schematic representation of the canonical boundary-value problem.}
	\label{fig:schem}
\end{figure}

In this section we present the formulation of the boundary-value problem, wherein the half space $z \leq 0$ is occupied by a HIDD material with complex refractive index $n_d$, and the half space $z \ge 0$ is occupied by a PMLID material.  A schematic representation of the boundary-value problem is provided in \fref{fig:schem}. 

\begin{figure}
	\begin{center}
		\begin{tabular}{c}
			(a) \includegraphics[width=0.8\linewidth]{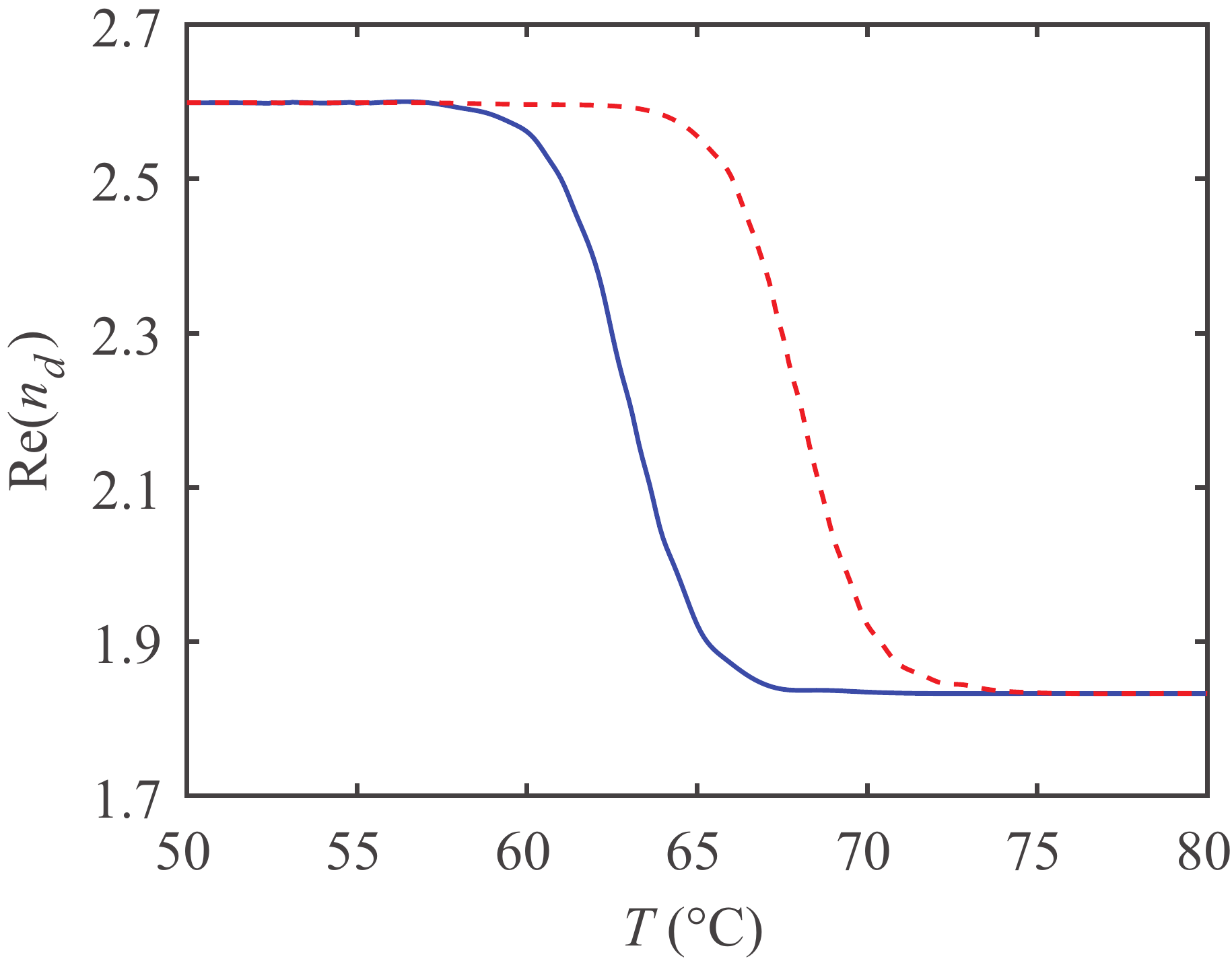}\\
			(b) \includegraphics[width=0.8\linewidth]{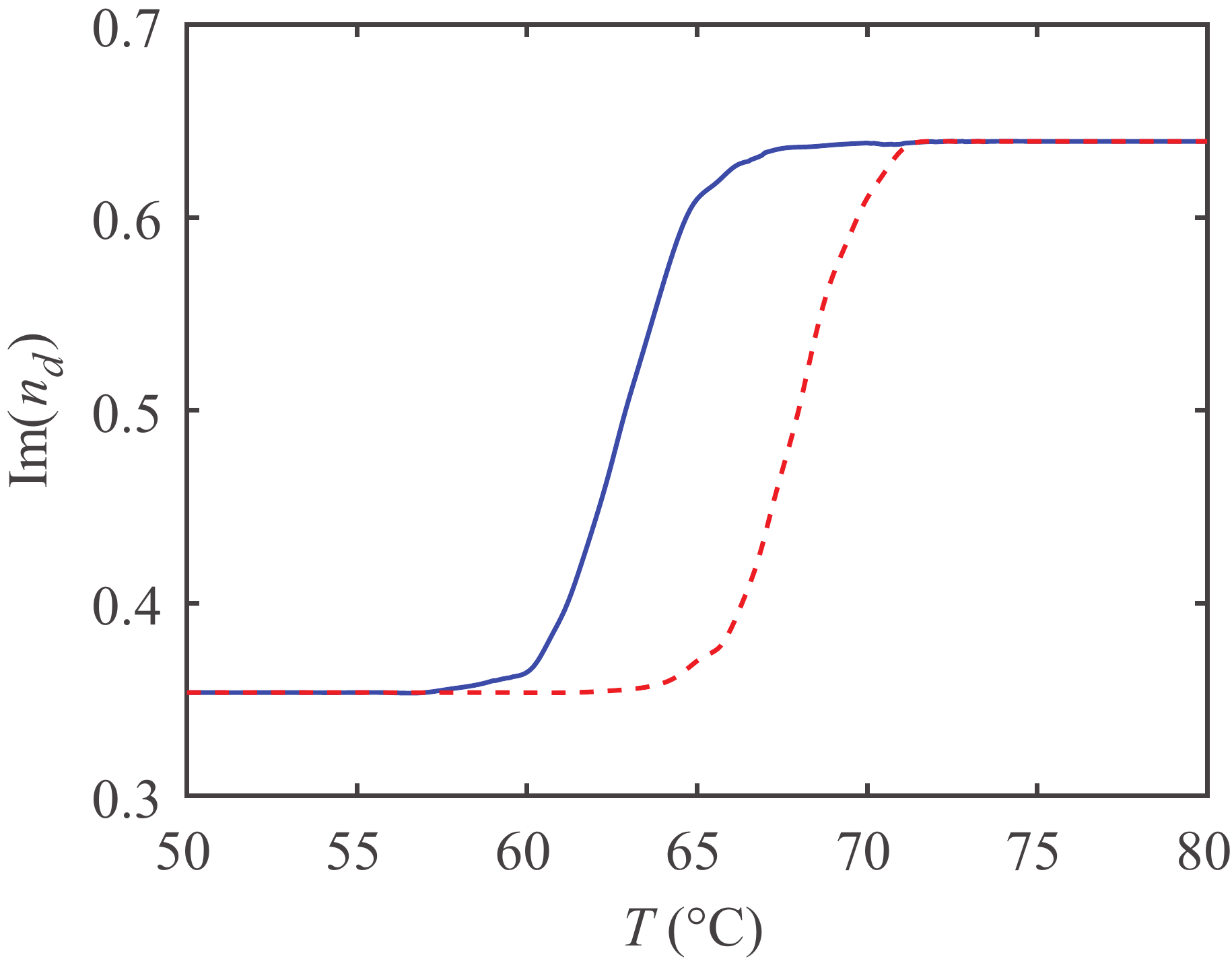}
		\end{tabular}
	\end{center}
	\caption{(Color online) Refractive index of VO$_2$ at  $\lambdao=800$~nm as a function of temperature  (in degree Celsius): (a) real part of $n_d$; (b) imaginary part of $n_d$. Solid (blue) curve: cooling phase; dashed (red) curve: heating phase.}
	\label{fig:VO2}
\end{figure}

The HIDD material was taken to be VO$_2$ whose refractive index $n_d$ in the range $\quadr{50,80}$\gra\ is a function of temperature,  with a thermal hysteresis feature.
 The thermal hysteresis of VO$_2$ has been quantified by Cormier \textit{et al.} for the 
 near-infrared   regime~\cite{Corm}, and is illustrated in Fig.~\ref{fig:VO2}. Specifically, 
$n_d$ in the range $\quadr{50,80}$\gra\ shows two different dependencies on temperature: one for heating the material from 50 to 80\gra\ and the other for cooling the material from 80 to 50\gra.
As shown in Fig.~\ref{fig:VO2}, graphs of  both the real and imaginary parts  of $n_d$ plotted against temperature exhibit the typical hysteresis shape
(as arises for plots of magnetization versus magnetic field in the case of  magnetic hysteresis \cite{hysteresis}, for example)
 with the 
heating graph (dashed red curve)  lying above the cooling graph (solid blue curve)  for the real part of $n_d$ and lying below for the imaginary part of $n_d$.

The unit cell of the PMLID material is constructed from  $9$ lossless dielectric layers. Each layer is an alloy of  silicon oxide and silicon nitride,  the proportion of these being different for  each layer~\cite{FaryadJOSAB}.
The sequence of the values of the refractive index $n_j$ $(j=1,...,9)$ in the unit cell of the PMLID material is  shown in \tref{tab:PMLID}, where the layer numbered $j = 1$    has the lowest value of the Cartesian coordinate $z$.
Each layer has a
thickness of  $75$~nm.

\begin{table}
	\caption{\bf Refractive index $ n_j $ of the dielectric layers $ j $ in the 
		unit cell of  the PMLID material at $\lambdao=800$~nm.
		\label{tab:PMLID}
	}
	\begin{center} 
		\begin{tabular}{r|c}
			$j$ & $n_j$  \\
			\hline
			$1$ & $ 1.9656$\\
			$2$ & $ 1.8062$\\
			$3$ & $ 1.7650$\\
			$4$ & $ 1.7243$\\
			$5$ & $ 1.6518$\\
			$6$ & $ 1.6105$\\
			$7$ & $ 1.5590$\\
			$8$ & $ 1.5174$\\
			$9$ & $ 1.4748$\\
			\hline
		\end{tabular}
	\end{center}
\end{table}

As both partnering  materials are isotropic, the canonical boundary-value problem can be partitioned into  two independent problems, one for $p$-polarized Tamm waves and the other for  $s$-polarized Tamm waves.  Let the Cartesian unit vectors be identified as $\ux$, $\uy$, and $\uz$.
Without loss of generality,  we consider  Tamm  waves propagating parallel to the unit vector $\ux$. The amplitude of any ESW is required to  decay as  $z \to \pm \infty$. With  $q$ as  the generally complex-valued wavenumber, the electric and magnetic field phasors can be represented everywhere as
\begin{equation}
\left.\begin{array}{l}
\#E(\#r)= \quadr{e_x(z)\ux+e_z(z)\uz}\exp({iqx})\\[5pt]
\#H(\#r)= h_y(z) \uy\exp({iqx})
\end{array}\right\}\,, \quad z \in(-\infty,\infty)\,,
\label{eq:EHpmlid}
\end{equation}
for the $p$-polarized Tamm waves, and as
\begin{equation}
\left.\begin{array}{l}
\#E(\#r)=  e_y(z) \uy\exp({iqx})\\[5pt]
\#H(\#r)= \quadr{h_x(z)\ux+h_z(z)\uz}
\exp({iqx})
\end{array}\right\}\,, \quad z \in(-\infty,\infty)\,,
\label{eq:EHpmlid-s}
\end{equation}
for the  $s$-polarized Tamm waves,
where $e_x(z)$, $e_z(z)$, and $h_y(z)$ and $e_y(z)$, $h_x(z)$, and $h_z(z)$ are unknown functions.
For the $p$- and $s$-polarized Tamm waves, labeled by the subscript $\ell \in \lec p,s \ric$, the dispersion equation which determines the wavenumber $q$ can be written as
\begin{equation}
\det\{\=M_{\,\ell}(q)\}=0\,, \qquad \quad  \ell \in \lec p, s \ric,
\label{detM}\end{equation}
where $\=M_{\,\ell}$ is a $2\times 2$ matrix  with functional dependence on $q$,
 whose derivation is described in detail elsewhere~\cite{ChJNP15}.
A numerical code was developed to extract  wavenumbers $q$  from \eqref{detM}.
The propagation distance~---~i.e., the distance along the direction of propagation $\ux$ over which the fields reduce by a factor of $\exp({-1})= 0.367$, and the power density reduces by a factor of  $\exp({-2}) = 0.135$~---~is calculated from the imaginary part of  $q$ as  $\propdist=1/{\rm Im}(q)$. We also define the normalized wavenumber 	$\tq=q/\ko$, where $\ko=2\pi/\lambdao$ is the free-space wavenumber.

\section{Numerical Results and Discussions}\label{sec:RaD}

At a  fixed free-space wavelength   $\lambdao=800$~nm,
\eqref{detM} was numerically solved with the temperature varying over the range   $\quadr{50,80}$\gra.   At each temperature, 
multiple solutions of \eqref{detM} exist. These solutions can be organized in branches with respect to temperature variation.

\begin{figure}[h!]
	\centering
	\includegraphics[width=0.95\linewidth]{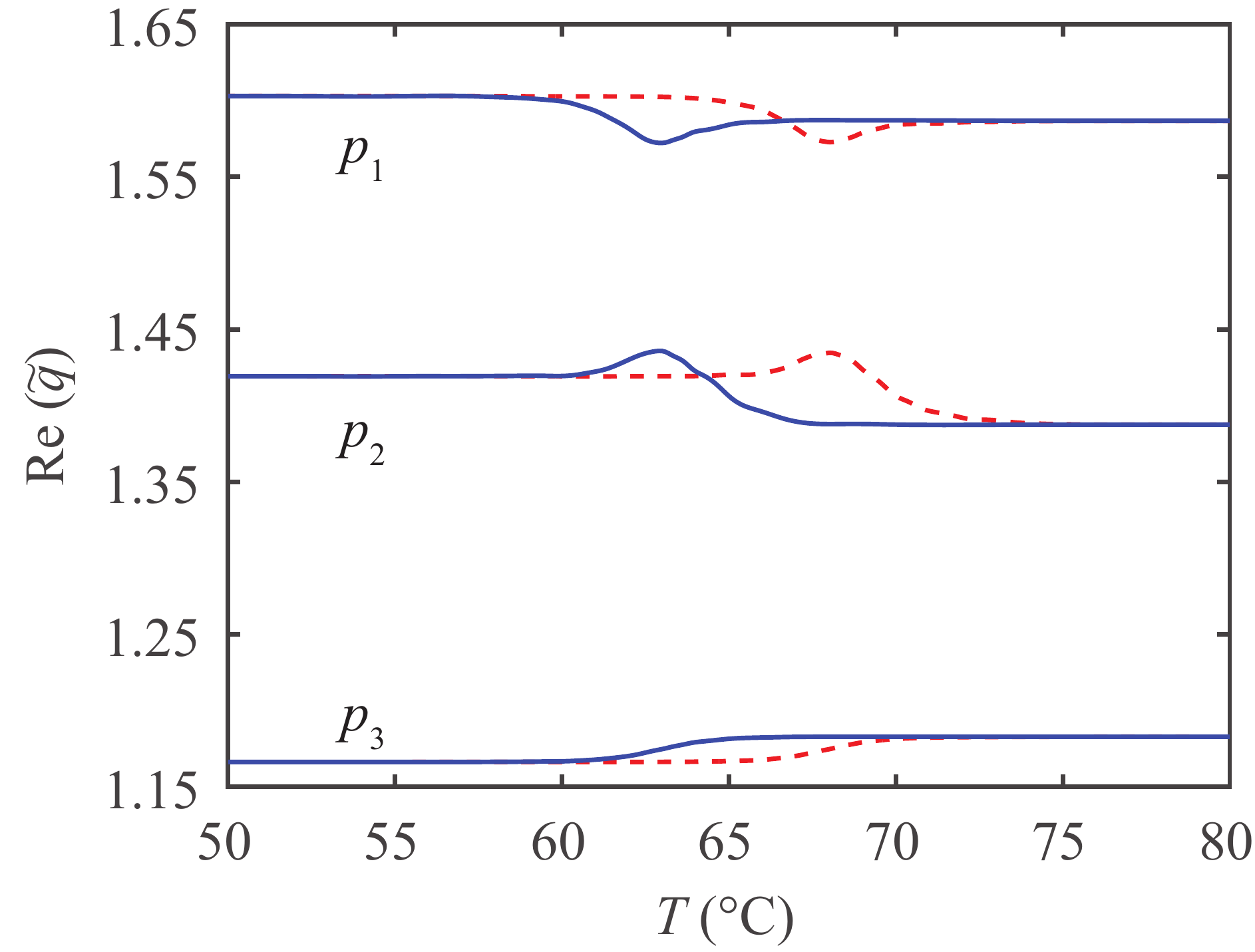}
	\caption[p-polarized, \req$>1$]{(Color online) \req\  versus temperature for $p$-polarized Tamm waves for which \req$>1$ at $T=50$\gra. Solid (blue) curve: cooling phase; dashed (red) curve: heating phase.}
	\label{fig:ppolreqmag1}
\end{figure}

\begin{figure}[h!]
	\centering
	\includegraphics[width=0.95\linewidth]{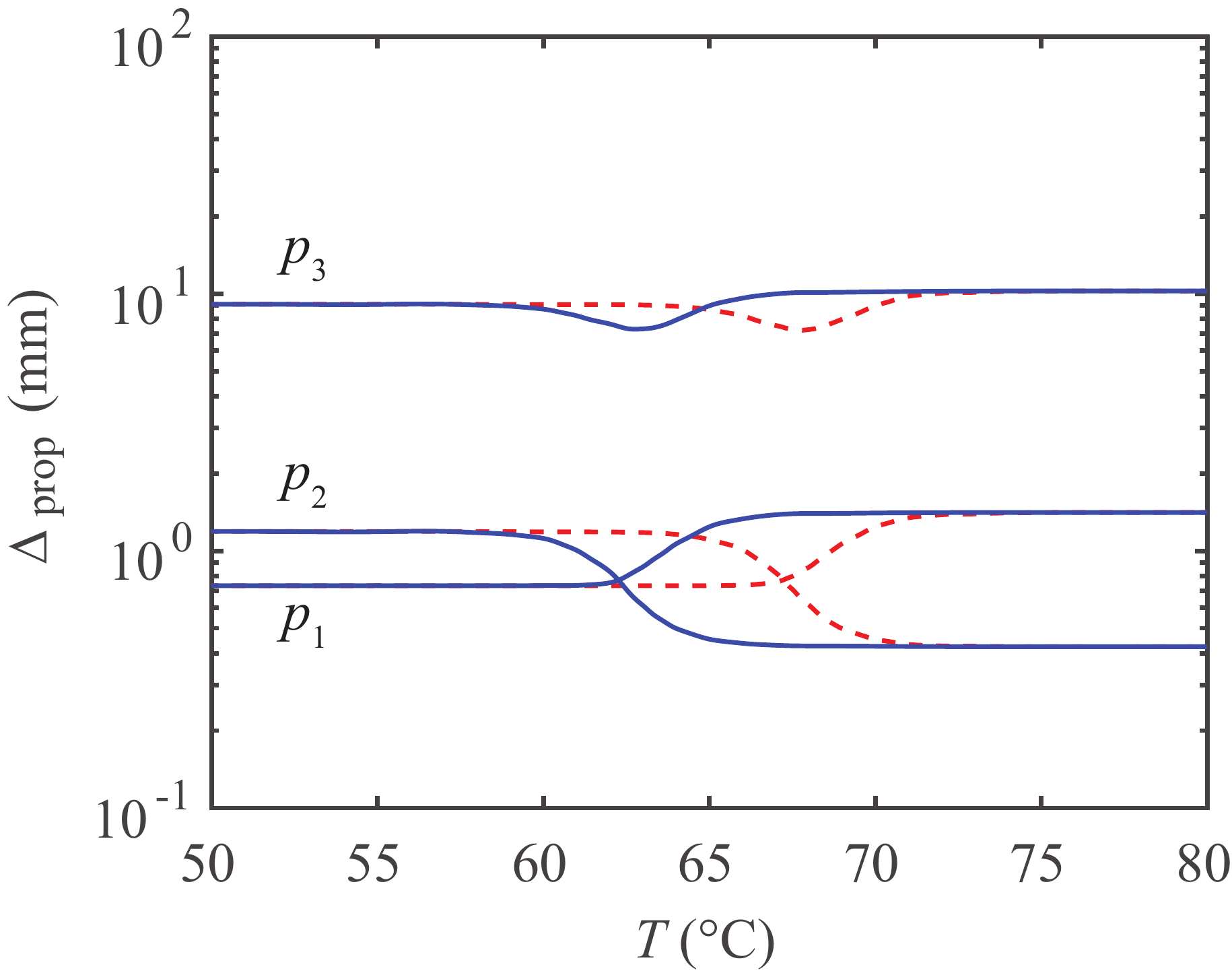}
	\caption{(Color online) $\propdist$ versus temperature for $p$-polarized Tamm waves for which \req$>1$ at $T=50$\gra.  Solid (blue) line: cooling phase; dashed (red) line: heating phase.}
	\label{fig:ppoldpropqmag1}
\end{figure}

\begin{figure}[h!]
	\centering
	\includegraphics[width=0.95\linewidth]{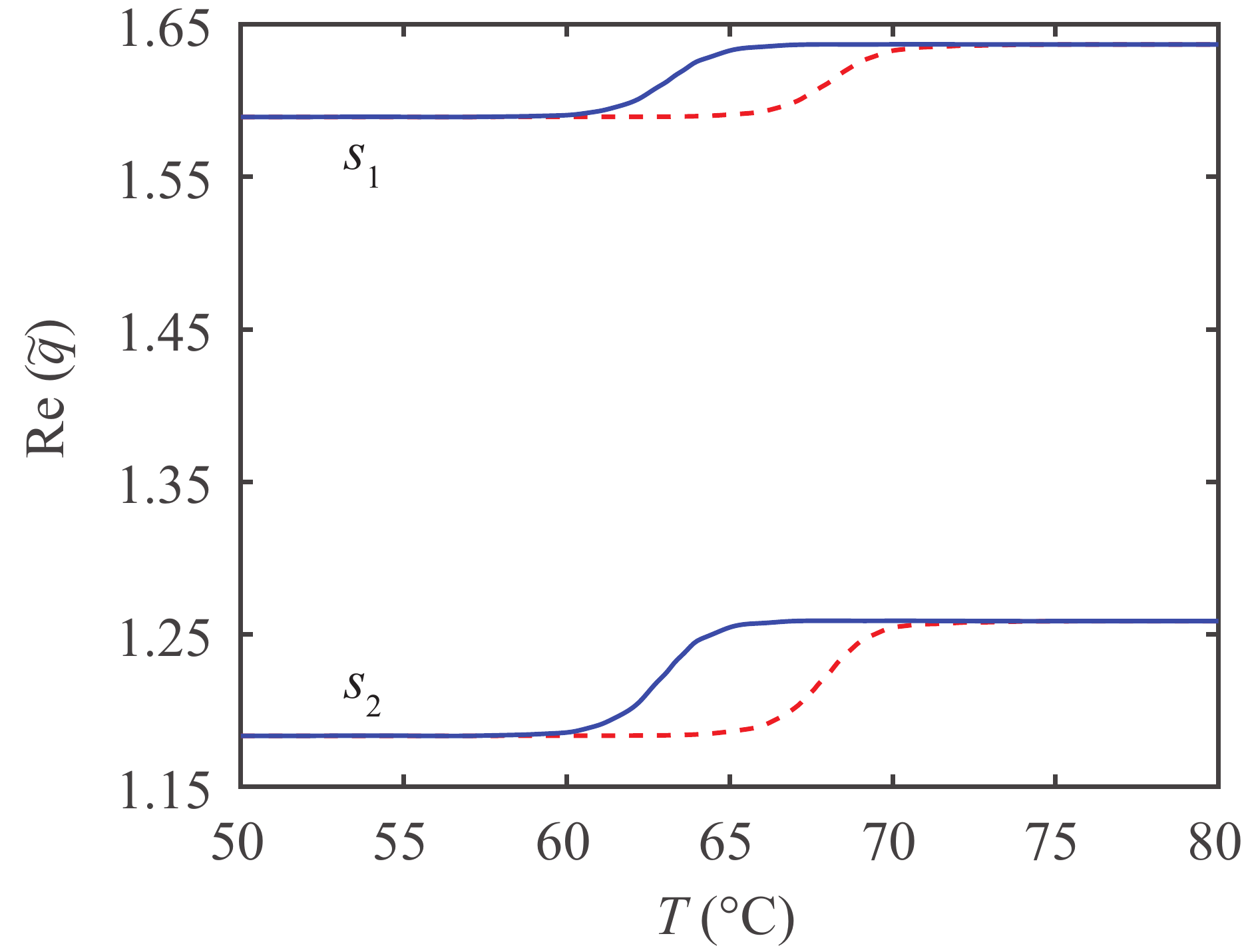}
	\caption[s-polarized, \req$>1$]{As Fig.~\ref{fig:ppolreqmag1} but for $s$-polarized Tamm waves.}
	\label{fig:spolreqmag1}
\end{figure}

\begin{figure}[h!]
	\centering
	\includegraphics[width=0.95\linewidth]{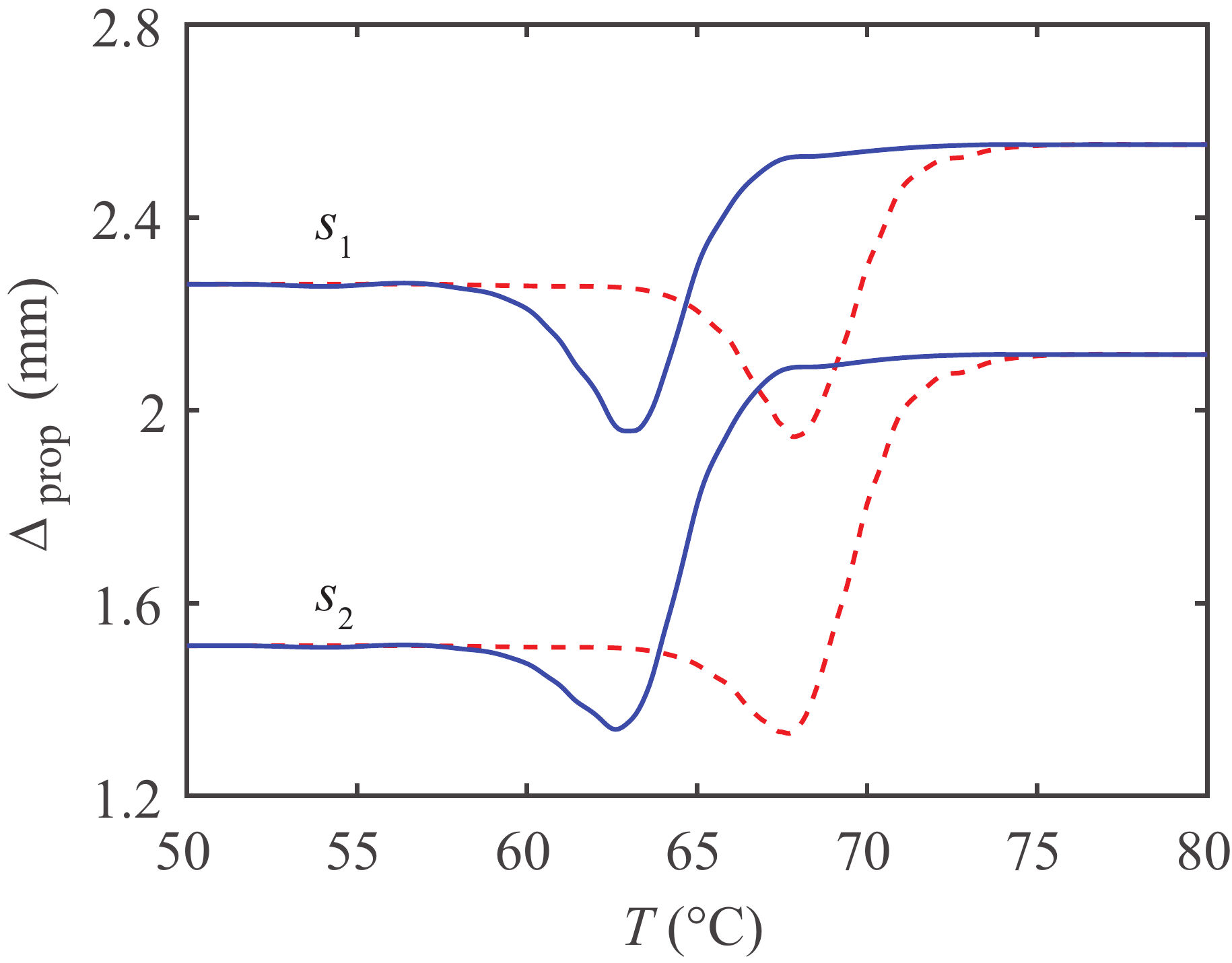}
	\caption{As Fig.~\ref{fig:ppoldpropqmag1} but for $s$-polarized Tamm waves.}
	\label{fig:spoldpropqmag1}
\end{figure}

\begin{figure}[h!]
	\centering
	\includegraphics[width=0.95\linewidth]{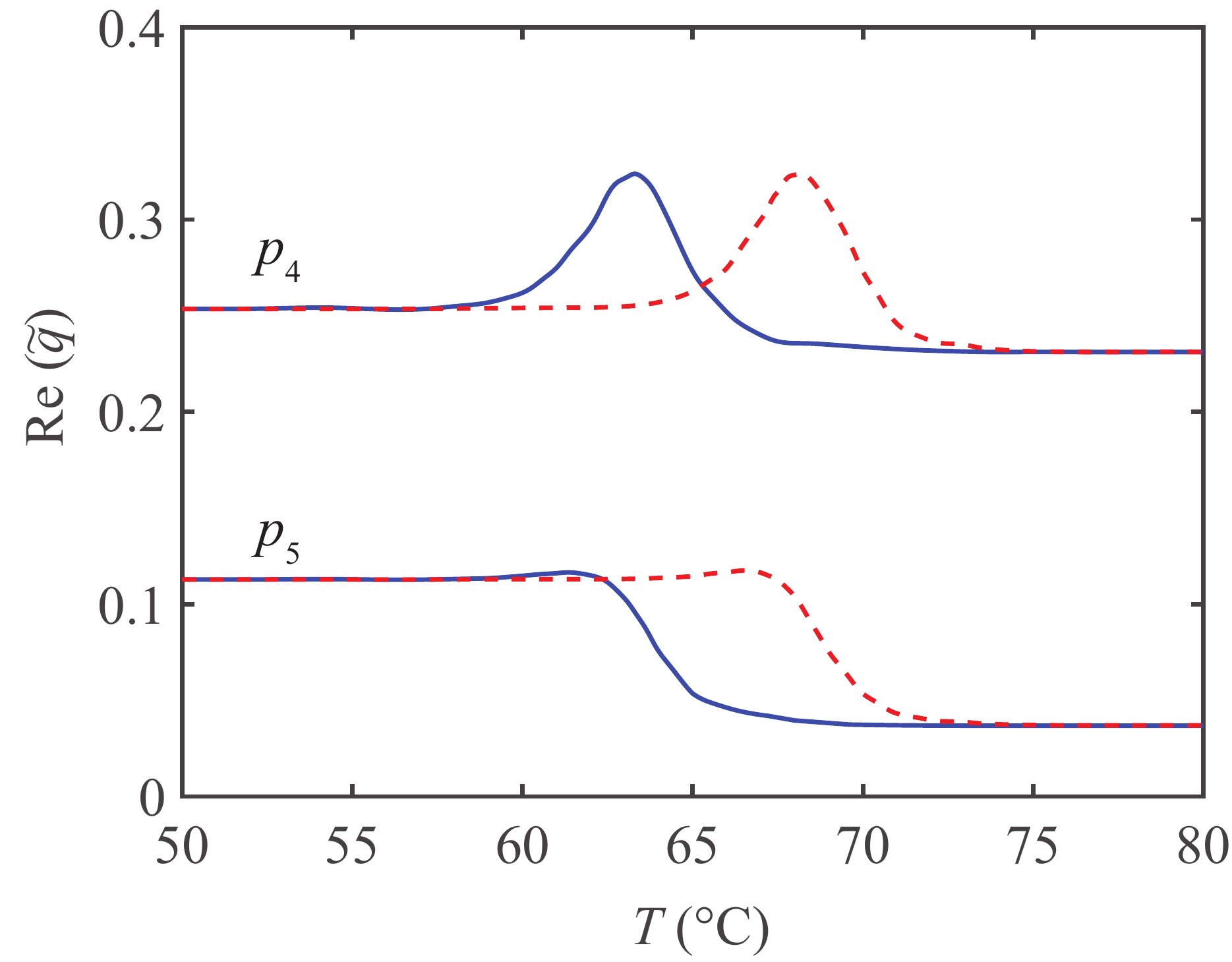}
	\caption{
	As Fig.~\ref{fig:ppolreqmag1} but for  $0.1<$\req$<1$ at $T=50$\gra. }
	\label{fig:Req_p_01-1}
\end{figure}

\begin{figure}[h!]
	\centering
	\includegraphics[width=0.95\linewidth]{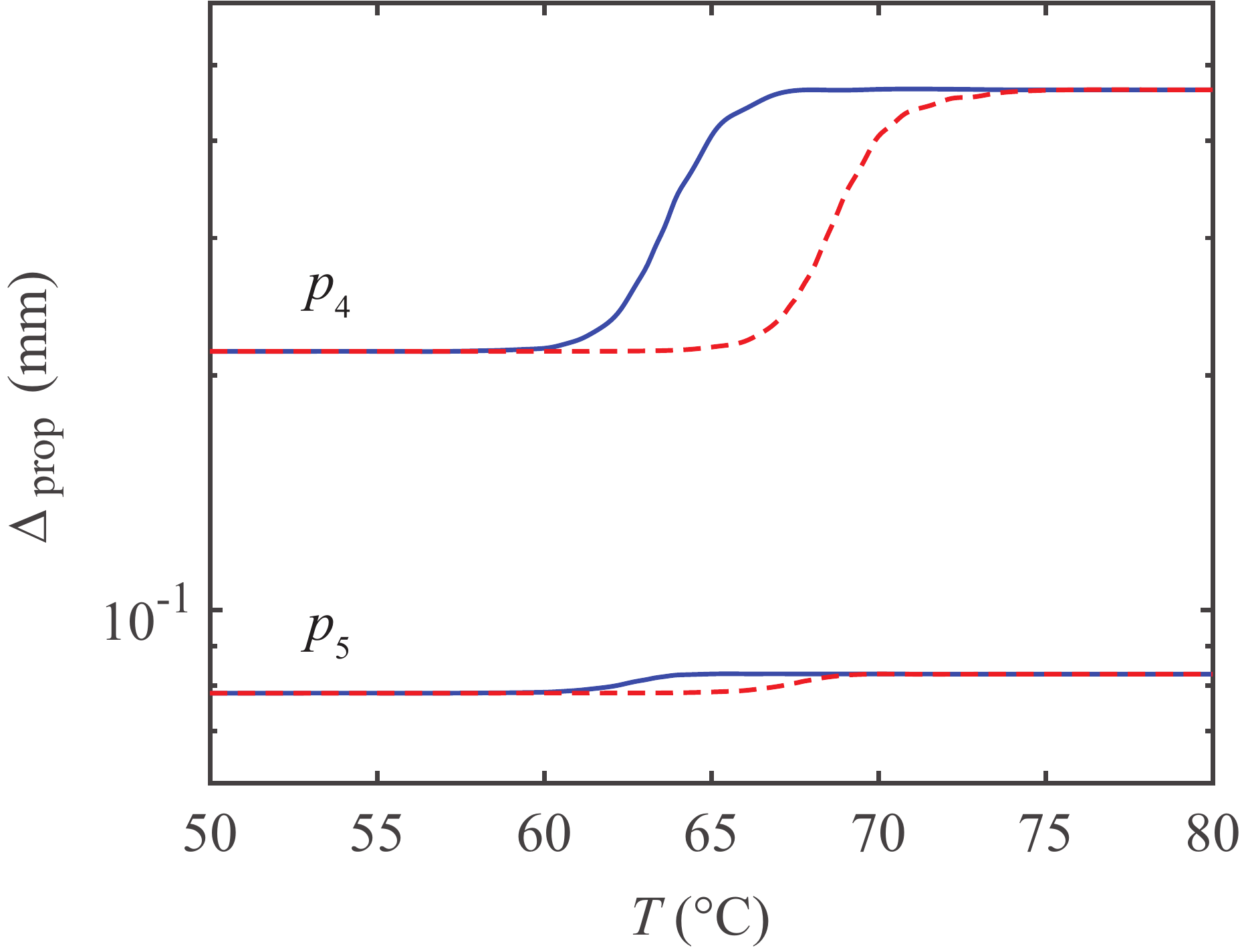}
	\caption{	As Fig.~\ref{fig:ppoldpropqmag1} but for  $0.1<$\req$<1$ at $T=50$\gra. }
	\label{fig:Dprop_p_01-1}
\end{figure}

\begin{figure}[h!]
	\centering
	\includegraphics[width=0.95\linewidth]{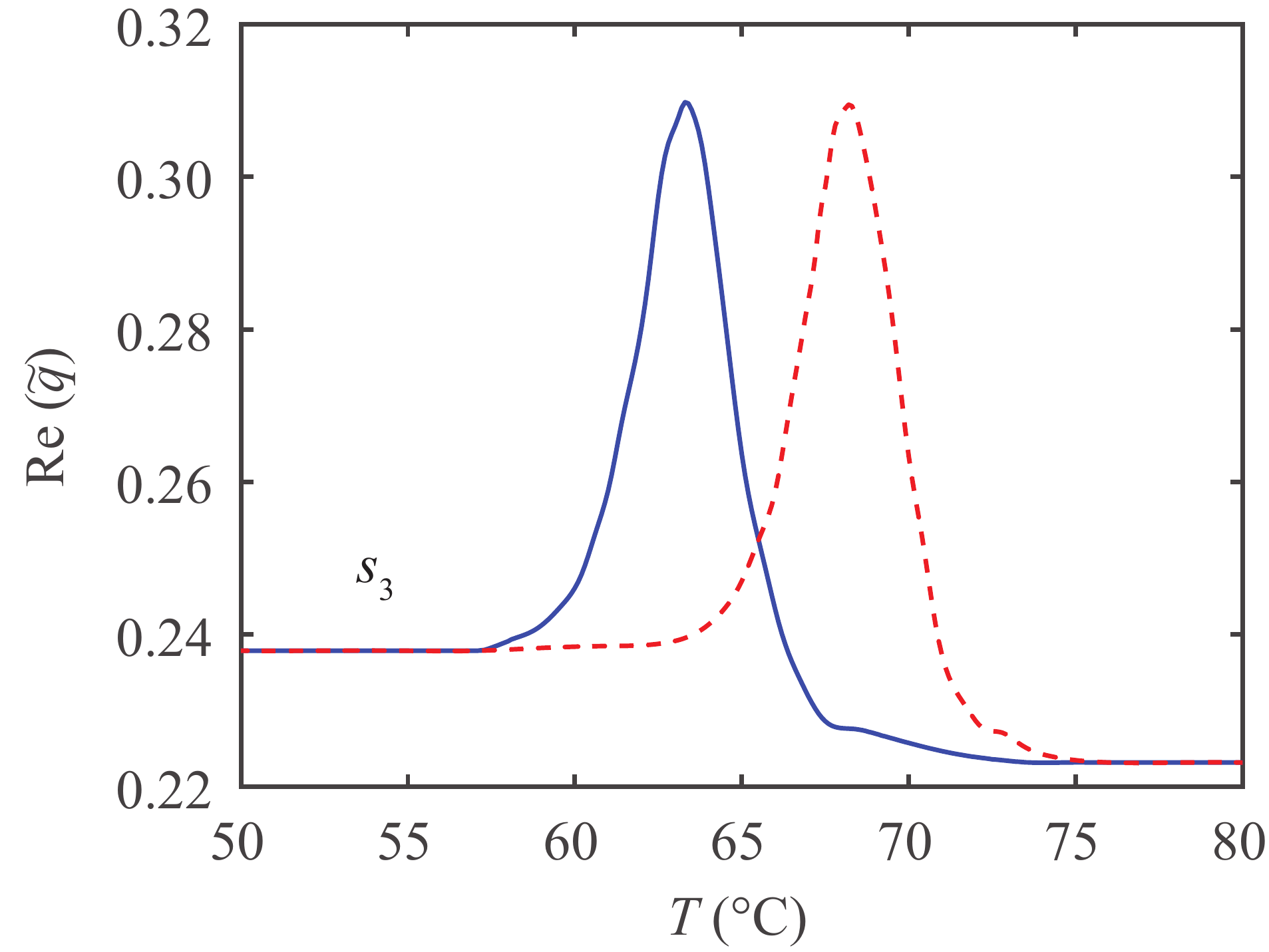}
	\caption{As Fig.~\ref{fig:spolreqmag1} but for  $0.1<$\req$<1$ at $T=50$\gra. 
	}
	\label{fig:Req_s_01-1}
\end{figure}

\begin{figure}[h!]
	\centering
	\includegraphics[width=0.95\linewidth]{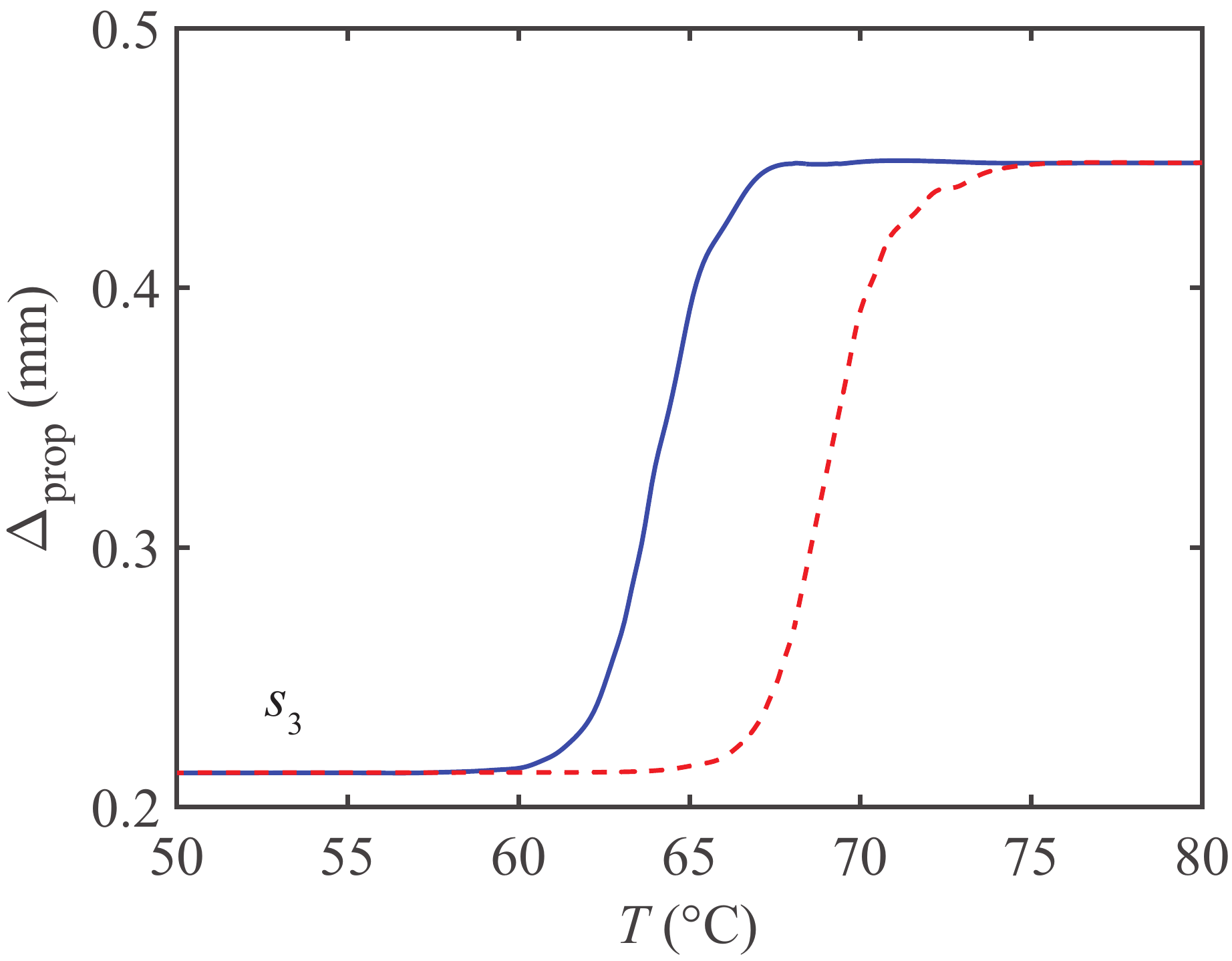}
	\caption{As Fig.~\ref{fig:spoldpropqmag1} but for  $0.1<$\req$<1$ at $T=50$\gra.}
	\label{fig:Dprop_s_01-1}
\end{figure}

\begin{figure}[h!]
	\centering
	\includegraphics[width=0.95\linewidth]{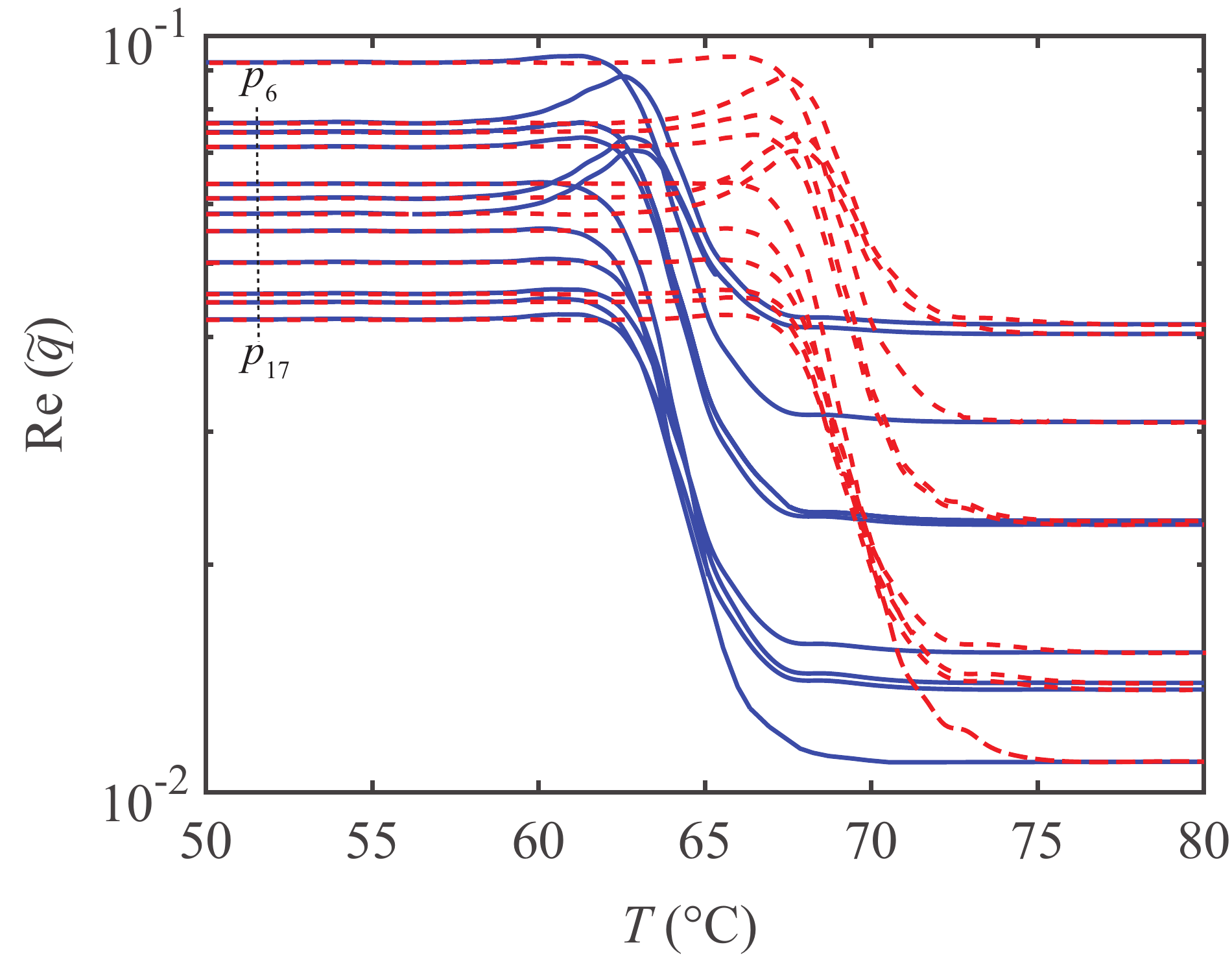}
	\caption{As Fig.~\ref{fig:ppolreqmag1} but for  $0.01<$\req$<0.1$  at $T=50$\gra. 
	}
	\label{fig:Req_p_001-01}
\end{figure}

\begin{figure}[h!]
	\centering
	\includegraphics[width=0.95\linewidth]{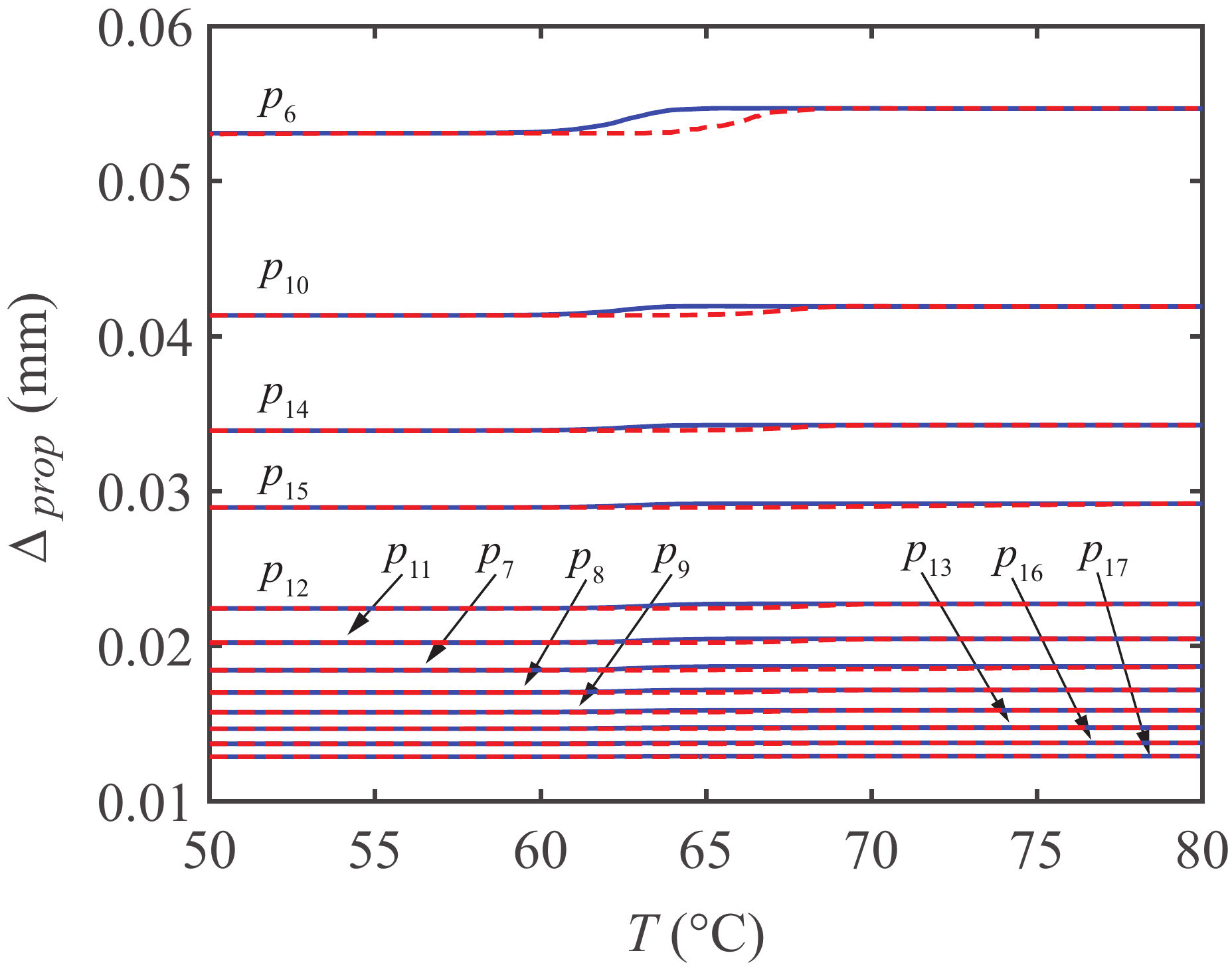}
	\caption{As Fig.~\ref{fig:ppoldpropqmag1} but for  $0.01<$\req$<0.1$  at $T=50$\gra. 
	}
	\label{fig:Dprop_p_001-01}
\end{figure}

\begin{figure}[h!]
	\centering
	\includegraphics[width=0.95\linewidth]{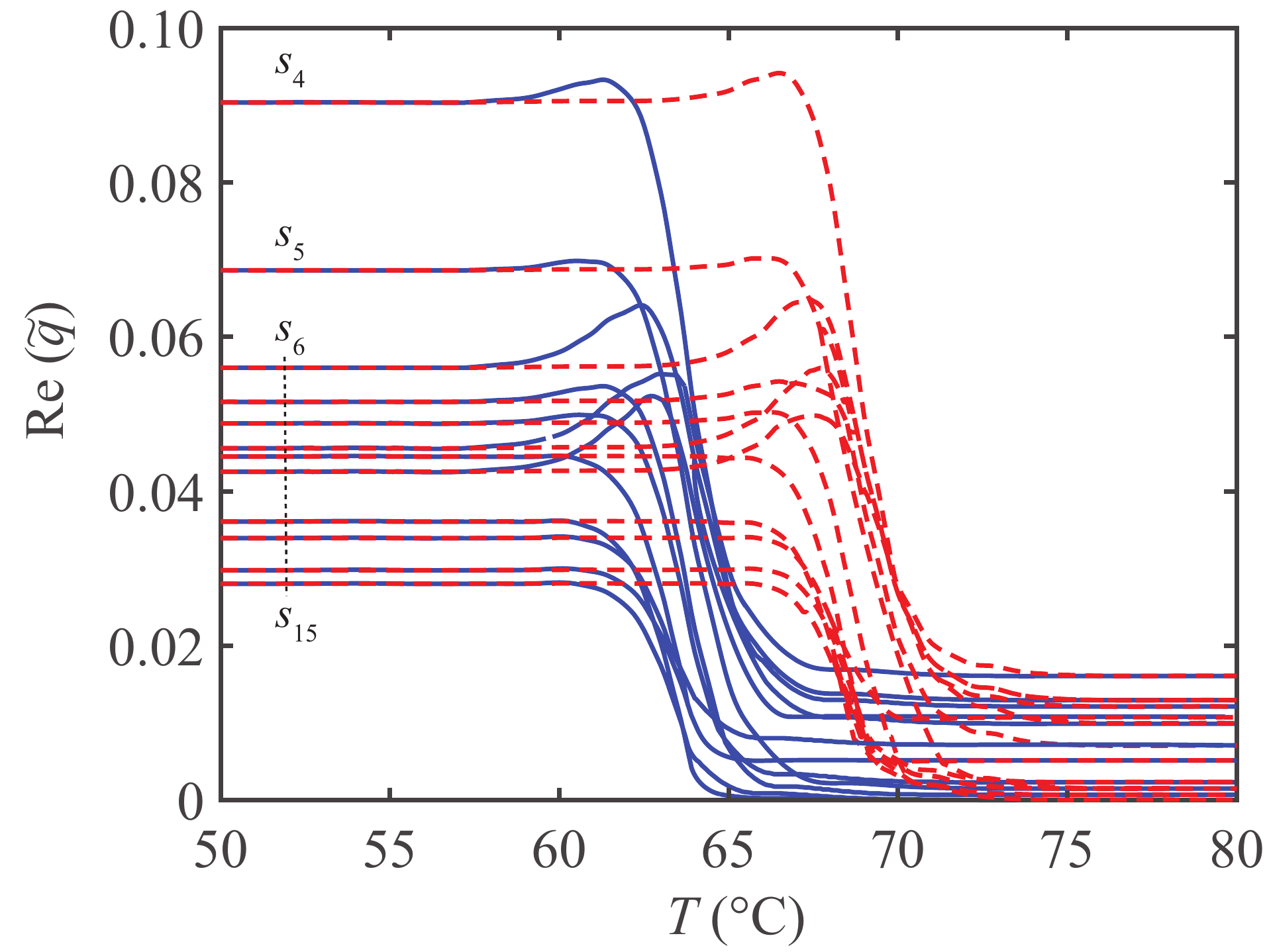}
	\caption{As Fig.~\ref{fig:spolreqmag1} but for  $0.01<$\req$<0.1$  at $T=50$\gra. 
	}
	\label{fig:Req_s_001-01}
\end{figure}

\begin{figure}[h!]
	\centering
	\includegraphics[width=0.95\linewidth]{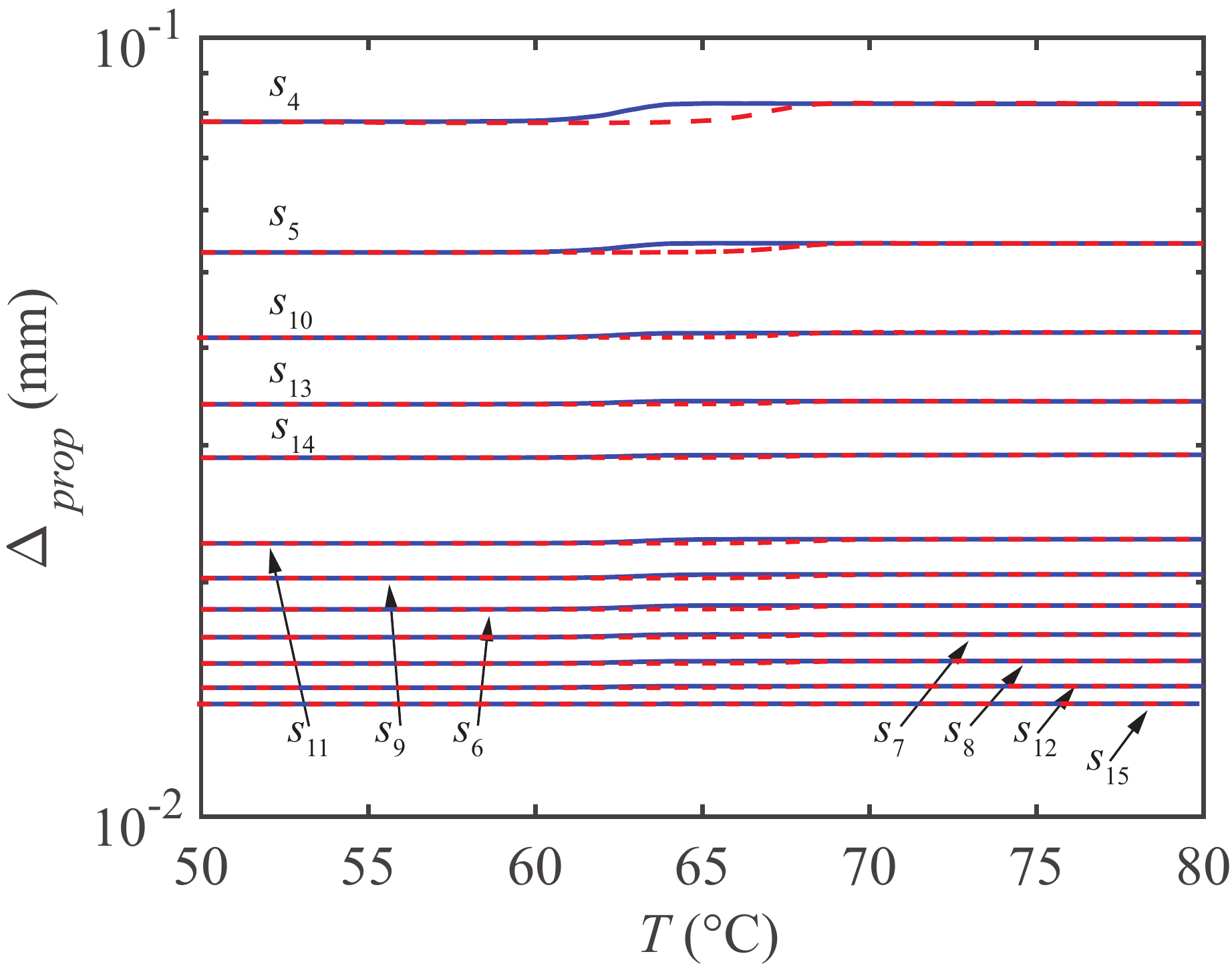}
	\caption{As Fig.~\ref{fig:spoldpropqmag1} but for  $0.01<$\req$<0.1$ at $T=50$\gra.
	}
	\label{fig:Dprop_s_001-01}
\end{figure}

Owing to the large number of solution branches,  it is convenient to present the solution branches in three groups  for each polarization state: 
\begin{itemize}
\item[(i)]
plots of Re$\le \tq \ri$ and $\propdist$ as the temperature increases from $50$\gra\ to $80$\gra\ (red dashed curves) and decreases from $80$\gra\ to $50$\gra\ (blue solid curves) for
solution branches with  \req$>1$ at $T=50$\gra\ are presented in Figs.~\ref{fig:ppolreqmag1} and \ref{fig:ppoldpropqmag1}  for $p$-polarized Tamm waves and in Figs.~\ref{fig:spolreqmag1} and \ref{fig:spoldpropqmag1}  for $s$-polarized Tamm waves;
 \item[(ii)] plots of Re$\le \tq \ri$ and $\propdist$ as the temperature increases from $50$\gra\ to $80$\gra\ (red dashed curves) and decreases from $80$\gra\ to $50$\gra\ (blue solid curves) for   solution branches with $0.1<$\req$<1$  at $T=50$\gra\
 are presented in Figs.~\ref{fig:Req_p_01-1} and \ref{fig:Dprop_p_01-1} for $p$-polarized Tamm waves and in Figs.~\ref{fig:Req_s_01-1} and \ref{fig:Dprop_s_01-1}  for $s$-polarized Tamm waves; and
 \item[(iii)]  plots of Re$\le \tq \ri$ and $\propdist$ as the temperature increases from $50$\gra\ to $80$\gra\ (red dashed curves) and decreases from $80$\gra\ to $50$\gra\ (blue solid curves) for  solution branches with $0.01<$\req$<0.1$  at $T=50$\gra\ are presented in Figs.~\ref{fig:Req_p_001-01} and \ref{fig:Dprop_p_001-01} for $p$-polarized Tamm waves and in Figs.~\ref{fig:Req_s_001-01} and \ref{fig:Dprop_s_001-01}  for $s$-polarized Tamm waves.
\end{itemize}

In each of Figs.~\ref{fig:ppolreqmag1}--\ref{fig:Dprop_s_001-01} there is a clear distinction between the heating and cooling phases. This distinction~---~which arises for both $p$-polarized and $s$-polarized Tamm waves~---~reflects hysteresis in the temperature dependence of the 
refractive index of VO$_2$  for the heating and cooling phases.   Each pair of  heating and cooling curves can be classified as either of signature \textsf{A} or of \textsf{B}, as   follows:

\begin{description}
\item[Signature \textsf{A}:] The extremum value  of \req\ is approximately the same in both heating and cooling phases, but it occurs at different temperatures in the two phases.  
Plots of $\propdist$ for the heating and cooling phases  exhibit the typical  hysteresis shape \cite{hysteresis}.  The hysteresis shape depends on whether  $\propdist$ 
  at $T=50$\gra\ is greater or smaller than its value at  $T=80$\gra.
Thus,  signature \textsf{A}  can be divided into two subsignatures:
\begin{description}
	\item[Subsignature \textsf{Aa}:]  $\propdist$ 
  at $T=50$\gra\ is  smaller than its value at  $T=80$\gra\;, and
		\item[Subsignature \textsf{Ab}:] $\propdist$ 
  at $T=50$\gra\ is  larger than its value at  $T=80$\gra.
	\end{description}

\item[Signature \textsf{B}:]  The extremum value of $\propdist$ is approximately the same in both heating and cooling phases, but occurs at different temperatures in the two phases.
Plots of \req\ for the heating and cooling phases  exhibit the typical  hysteresis shape \cite{hysteresis}.
For all solutions branches of signature \textsf{B},  \req\ 
  at $T=50$\gra\ is  smaller than its value at  $T=80$\gra.

\end{description}
In order to illustrate the foregoing classification, representative hysteresis shapes for  the  signatures  \textsf{Aa},   \textsf{Ab}, and  \textsf{B} are identified in \tref{tab:types}.

\begin{table}[h!]\setlength\tabcolsep{4pt}
	\centering
	\caption{\bf Representative  hysteresis shapes.
		\label{tab:types}}
	\begin{tabular}{>{\centering\arraybackslash}m{15mm}| >{\centering\arraybackslash}m{27mm}| >{\centering\arraybackslash}m{27mm}}
		\textbf{Solution signature} & \req & $\propdist$\\ \hline
		\textsf{Aa} &no typical hysteresis shape&\vspace{2mm} \includegraphics[height=20mm]{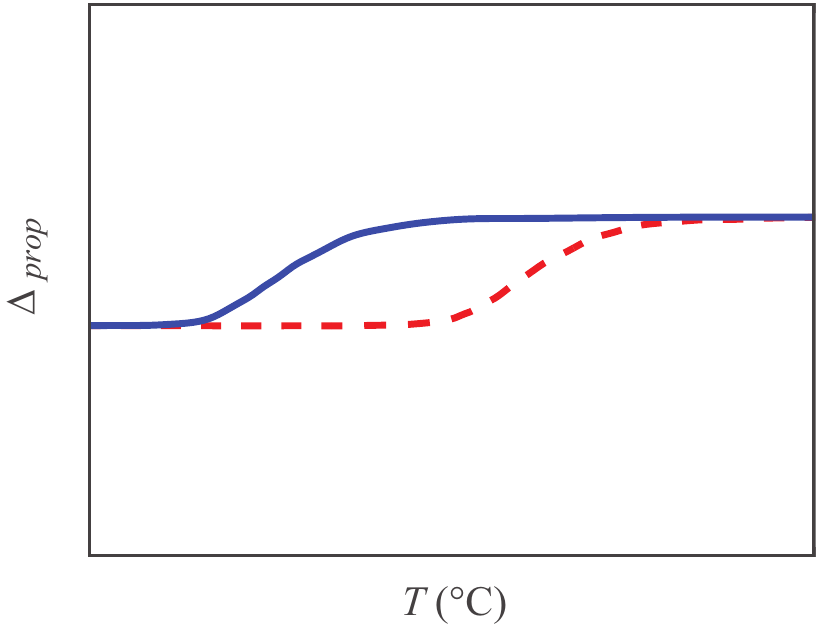}\\ \hline
		\textsf{Ab} &no typical hysteresis shape&\vspace{2mm}  \includegraphics[height=2cm]{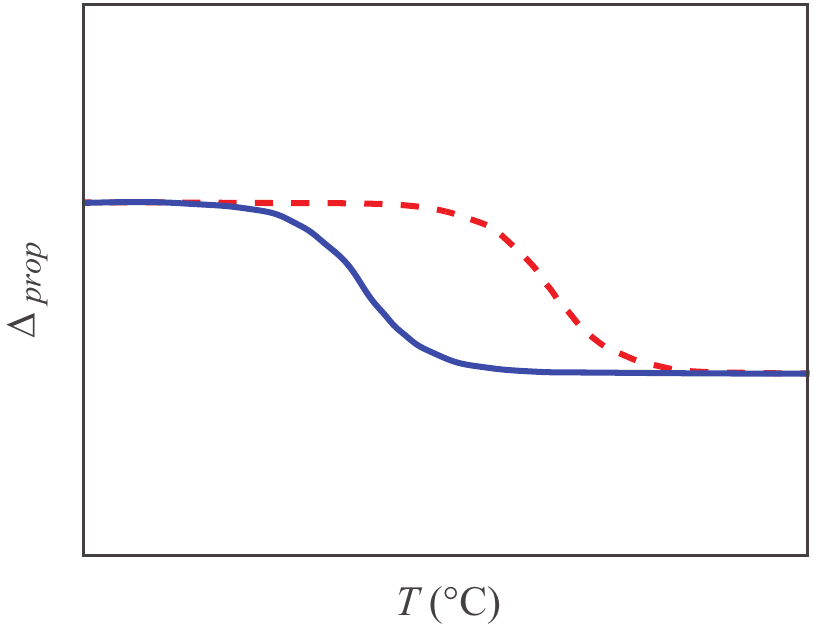}\\ \hline
		\textsf{B} &\vspace{2mm} \includegraphics[height=2cm]{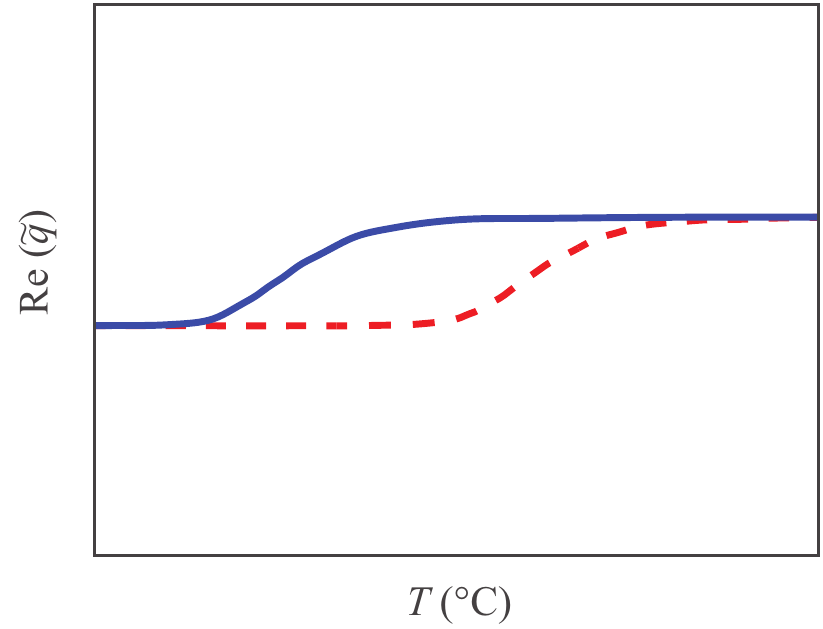} & no typical hysteresis shape \\ \hline

	\end{tabular}
\end{table}

Figures~\ref{fig:ppolreqmag1} and \ref{fig:ppoldpropqmag1} illustrate  \req\ and $\propdist$, respectively,  for  $p$-polarized Tamm waves when \req$>1$. The solutions are organized in three branches labeled $p_1$ to $p_3$. The branch $p_1$ is of signature \textsf{Aa}, branch $p_2$ is of signature \textsf{Ab}, while branch $p_3$ is of signature \textsf{B}. Figures~\ref{fig:spolreqmag1} and \ref{fig:spoldpropqmag1} depict \req\ and $\propdist$, respectively, for    $s$-polarized Tamm waves when \req$>1$. In these figures the solutions are organized in two branches labeled $s_1$ and $s_2$. Both branches are of signature \textsf{B}.


Figures~\ref{fig:Req_p_01-1} and \ref{fig:Dprop_p_01-1} illustrate  \req\ and $\propdist$, respectively,  for   $p$-polarized Tamm waves when \req\ at $T=50$\gra\  takes a value greater than $0.1$ and less than $1$. The solutions are organized in two branches labeled $p_4$ and $p_5$. Both branches are of signature \textsf{Aa}. Figures~\ref{fig:Req_s_01-1} and \ref{fig:Dprop_s_01-1} depict $\tq$ and $\propdist$ for  $s$-polarized Tamm waves when \req\ at $T=50$\gra\  takes a value between $0.1$ and $1$. The solitary solution represented in these figures, labeled $s_3$, is of signature \textsf{Aa}.

Figures~\ref{fig:Req_p_001-01} and \ref{fig:Dprop_p_001-01}  illustrate  \req\  and $\propdist$, respectively,  for   $p$-polarized Tamm waves when \req\ at $T=50$\gra\ takes a value between $0.01$ and  $0.1$. The solutions are organized in twelve branches labeled from $p_6$ to $p_{17}$ in descending order with respect to the \req\ values at $T=50$\gra. Solution branches in Fig.~\ref{fig:Req_p_001-01} have  rather similar \req\ values  and their graphs cross each other. In contrast, the $\propdist$ graphs of the $p_6-p_{17}$  solution branches do not intersect each other. The typical hysteresis shape  can be  recognized for most of  the $\propdist$ graphs. All solution branches from $p_6$ to $p_{17}$ are of signature \textsf{Aa}. Figures~\ref{fig:Req_s_001-01} and \ref{fig:Dprop_s_001-01} depict $\tq$ and $\propdist$, respectively, for the  $s$-polarization state when \req\ at $T=50$\gra\  takes a value between $0.01$ and $0.1$. The solutions are organized in twelve branches labeled from $s_4$ to $s_{15}$ in descending order with respect to the \req\ values at $T=50$\gra. All the remarks made for the $p$-polarization state represented in Figs.~\ref{fig:Req_p_001-01} and \ref{fig:Dprop_p_001-01}  also hold for the $s$-polarization state represented in Figs.~\ref{fig:Req_s_001-01} and \ref{fig:Dprop_s_001-01}. Thus, all branch solutions from $s_5$ to $s_{15}$ are of signature \textsf{Aa}.

When \req$<0.01$ at  $T=50$\gra, for both polarization states, a multitude of solutions were found with \req\ values that are very small and  very close to each other. These solutions are not presented graphically here as the corresponding  graphs are practically indistinguishable from each other.  It is worth noting that for solutions with very small \req\ values, the values of \imq\ are very high which results in tiny values of  $\propdist$. The corresponding Tamm  waves can therefore be identified as local ESWs. Finding these solutions is numerically challenging because extremely high resolution is needed. As a consequence, for \req$<10^{-4}$ our code is not able to reliably determine solutions.

\section{Concluding remarks}\label{sec:cr}
We numerically solved the boundary-value problem for Tamm  waves  (which may  be alternatively classified as  Uller--Zenneck waves)
guided by the interface of a homogeneous  isotropic dissipative dielectric material 
and a  periodic multilayered
isotropic dielectric material.
The HIDD material was chosen to have a temperature-dependent refractive index with a hysteresis feature.
A numerical code was developed to extract solutions of the  dispersion equation at a fixed wavelength for Tamm waves of both $p$- and $s$-polarization states.
A multitude of Tamm waves exist of both $p$- and $s$-polarization states which
show a  clear distinction between  heating and cooling phases,
reflecting the temperature-dependence of the refractive index  of the HIDD material for the two phases.
  Thus, the signatures of thermal hysteresis in Tamm-wave propagation were revealed. \blue{Parenthetically, the physical significance and applicability of our study are
bolstered by the fact that  experimental data has been used for  the relative
permittivity of VO${}_2$ as well as of  SiO${}_2$--SiN${}_2$ alloys of nine different compositions. }

\vspace{5mm}

\noindent {\bf Acknowledgment.}
  AL thanks the Charles Godfrey Binder Endowment at Penn State for ongoing support of his research.

\end{document}